\documentclass[aip,jap,preprint,superscriptaddress,amssymb,showpacs,showkeys,floatfix]{revtex4-1}

\usepackage{graphicx,amsmath,bm,xcolor}
\usepackage[version=3]{mhchem}
\usepackage{xspace}
\usepackage{subcaption}

\newcommand{\al}{\ensuremath{\alpha_2}\xspace}
\newcommand{\bo}{\ensuremath{\beta_\mathrm{o}}\xspace}
\newcommand{\g}{\ensuremath{\gamma}\xspace}
\newcommand{\uu}[1]{\ensuremath\,\mathrm{#1}}
\newcommand{\figWidth}{8.5cm}

\newcommand{\change}[1]{{\color{purple}#1}}
\newcommand{\Change}[1]{{\color{blue}#1}}

\begin{document}

\date{\today}
\title{Preferential site occupancy of alloying elements in TiAl-based phases}

\author{David Holec}
\email{david.holec@unileoben.ac.at}
\affiliation{Department of Physical Metallurgy and Materials Testing, Montanuniversit\"at Leoben, Franz-Josef-Strasse 18, A-8700 Leoben, Austria}

\author{Rajeev K. Reddy}
\affiliation{Department of Physical Metallurgy and Materials Testing, Montanuniversit\"at Leoben, Franz-Josef-Strasse 18, A-8700 Leoben, Austria}

\author{Thomas Klein}
\affiliation{Department of Physical Metallurgy and Materials Testing, Montanuniversit\"at Leoben, Franz-Josef-Strasse 18, A-8700 Leoben, Austria}


\author{Helmut Clemens}
\affiliation{Department of Physical Metallurgy and Materials Testing, Montanuniversit\"at Leoben, Franz-Josef-Strasse 18, A-8700 Leoben, Austria}

\begin{abstract}
First principles calculations are used to study the preferential occupation of ternary alloying additions into the binary Ti-Al phases, namely \g-TiAl, \al-\ce{Ti3Al}, \bo-TiAl, and B19-TiAl. While the early transition metals (TMs, group IVB, VB , and VIB elements) prefer to substitute for Ti atoms in the \g-, \al-, and B19-phases, they preferentially occupy Al sites in the \bo-TiAl. Si is in this context an anomaly, as it prefers to sit on the Al sublattice for all four phases. B and C are shown to prefer octahedral Ti-rich interstitial positions instead of substitutional incorporation. The site preference energy is linked with the alloying-induced changes of energy of formation, hence alloying-related (de)stabilisation of the phases. We further show that the phase-stabilisation effect of early TMs on \bo-phase has a different origin depending on their valency. Finally, an extensive comparison of our predictions with available theoretical and experimental data (which is, however, limited mostly to the \g-phase) shows a consistent picture.
\end{abstract}



\maketitle

\section{Introduction}

First principles methods have become an important part of the cutting edge materials research. One of a particularly important area is high throughput searches focused on scanning a large amount of systems, with the aim to further improve current or find novel materials \cite{Curtarolo2013-sp}, for example by establishing alloying-related trends for knowledge-based materials design \cite[e.g.,][]{Holec2013-yt}. When constructing an atomistic model for a solid solution based on an ordered multi-element compound \change{(as needed for, e.g., the sublattice approach used in the thermodynamic CALPHAD assessment of phase diagrams\cite{Zhang2015-ae})}, a question arises whether an alloying element $X$ prefers to substitute for a specific element forming the parent compound, or whether it is more likely to be incorporated into interstitial positions.

\begin{figure}[b]
  \setlength{\unitlength}{1cm}
  \begin{picture}(8,8.2)(0,-0.2)
    \put(0,0){\includegraphics[width=\figWidth]{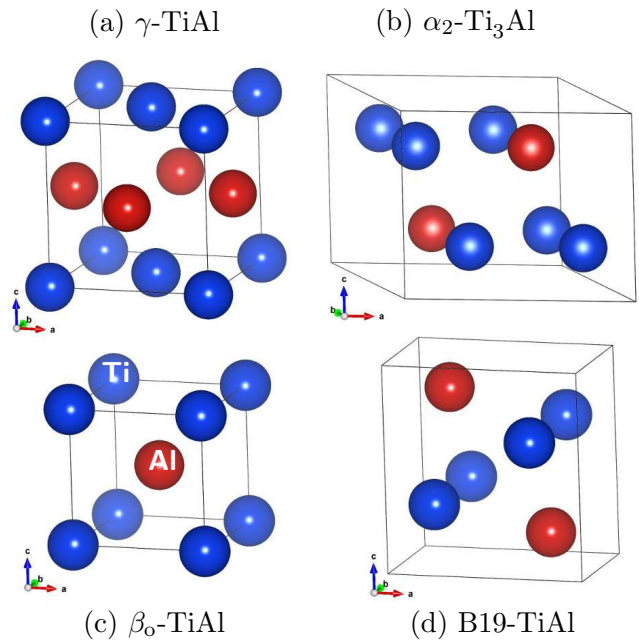}}
    \put(0,7.8){\parbox{4cm}{\centering (a) \g-TiAl}}
    \put(4,7.8){\parbox{4cm}{\centering (b) \al-\ce{Ti3Al}}}
    \put(0,-0.2){\parbox{4cm}{\centering (c) \bo-TiAl}}
    \put(4.5,-0.2){\parbox{4cm}{\centering (d) B19-TiAl}}
    \put(1.3,3.13){\mbox{\color{white}\sffamily\bfseries Ti}}
    \put(1.92,1.95){\mbox{\color{white}\sffamily\bfseries Al}}
  \end{picture}
  \caption{Ti-Al phases considered in the present study: (a)~\g-TiAl, (b)~\al-\ce{Ti3Al}, (c)~\bo-TiAl, and (d)~B19-TiAl. Blue and red atoms represent Ti and Al atoms, respectively. Structures were visualised using the VESTA3 software package~\cite{VESTA}.}\label{fig:structures}
\end{figure}

In this paper we focus on the Ti-Al intermetallic system \cite{Appel2011-tf, Clemens2013-jd}, a basis of modern light-weight alloys with high strength, creep and oxidation resistance. The existing literature dealing with atomistic modelling of TiAl alloys considers almost exclusively the \g-TiAl (L$1_0$ structure which is a tetragonally deformed fcc lattice, space group $P4/mmm$, Fig.~\ref{fig:structures}a), a ground state structure of the stoichiometric $\ce{Ti}_{0.5}\ce{Al}_{0.5}$ phase. However, microstructure of the state-of-the-art TiAl alloys consist of more phases, namely hexagonal \al-\ce{Ti3Al} (D$0_{19}$, space group $P6_3/mmc$, Fig.~\ref{fig:structures}b), cubic \bo-\ce{TiAl} (B2, space group $Pm\bar3m$, Fig.~\ref{fig:structures}c) or orthorhombic B19-\ce{TiAl} (space group $Pmma$, Fig.~\ref{fig:structures}d), to name a few~\cite{Appel2011-tf}.

Additionally, many of the previous investigations considered impact of a single or only a few alloying elements on the overall alloy properties. For example, \citet{Kim2000-hv} predicted that Ru and Pd preferentially substitute for Al in \g-TiAl. \citet{Dang2007-ct} concluded that also V, Cr, Mn, Ni, Fe, and Co preferentially occupy Al sites in \g-TiAl. V, Cr, Mn, and B were also treated by \cite{Khowash1993-gy}. They concluded that these elements favour Ti sites in \g-TiAl, which is in a direct disagreement with \citet{Dang2007-ct}. Other works reported on Mn and Ni \cite{Bauer1996-vc}, Nb, V, Cr, and Mn \cite{Erschbaumer1993-ih, Wolf1996-ms}, Mg, V, Cr, Mn, Ga, and Mo \cite{Jinlong1992-pg}, Si, Nb, Mo, Ta, and W \cite{Woodward1998-wl}, and Nb, Mo, Ni, and Ag \cite{Zhou2010-ev}. The most extensive calculations were done by \citeauthor{Song2000-oz}~\cite{Song2000-oz,Song2002-cy} \Change{and \citet{Jiang2008-sn}} and include most of the 3$d$ and some 4$d$ elements \Change{(\citet{Jiang2008-sn} discussed also 5$d$ elements)}. All these reports, however, consider the site preference only in the \g-TiAl phase. While \citet{Singh2008-ek} and \citet{Holec2015-gs} discussed the site preference for Mo in the \bo-phase, predictions for other elements in the \bo-phase are missing. Site preference for some TM elements in the \al-phase was reported by \citet{Hao1999-ex} and \citet{Benedek2005-oq}. However, predictions for other phases appearing in the equilibrium TiAl(+$X$) phase diagram and/or elements of interest (e.g., Si)\change{, which are a basis of novel alloying concepts,} are still missing. \change{It should be mentioned that while the information on the sublattice occupation is crucial for knowledge-based materials development, the possible experimental approaches are limited to only a few techniques, e.g., atom probe tomography, and even when successfully applied they can hardly every cover a vast range of elements and phases as done in the present study.}

Therefore, in this paper we employ first principles calculations to investigate site preference and alloying-induced changes in the phase stability for B, C, Si, V, Cr, Zr, Nb, Mo, Hf, Ta, and W for the \g-, \al-, \bo-, as well as B19-phase. The selection of the elements reflects the typical alloying elements used in modern TiAl-based alloys~\cite{Appel2011-tf, Clemens2013-jd}. All ternary additions are treated as substitutional defects in the first place, but B and C will be also discussed in the view of incorporation on interstitial sites.

\section{Methods}

The quantum mechanical calculations were performed in the framework of Density Functional Theory (DFT) \cite{Hohenberg1964-in, Kohn1965-rd} as implemented in the Vienna Ab initio Simulation Package (VASP) \cite{Kresse1996-tg,Kresse1996-gt}. Projector augmented-wave pseudopotentials \cite{Kresse1999-if} together with the generalised gradient approximation (GGA) parametrised by \citet{Wang1991-ca} for the exchange and correlation (xc) effects were used to describe the interactions between electrons and ions on the quantum level. Some test calculations with the xc potential within the local density approximation (LDA) \cite{Hohenberg1964-in} were also performed. With plane-wave energy cut-off energy of $400\uu{eV}$ and more than $19\,000$ $k$-points$\times$atoms, the total energy is converged to an accuracy of about $1\uu{meV/at.}$ All calculations were spin-nonpolarised.

The basic structural building blocks, i.e., a conventional cubic cell for the \g-TiAl phase, and unit cells for the other phases, are shown in Fig.~\ref{fig:structures}. The site preference and stability of solid solutions were evaluated using supercells containing 32 atoms, i.e., $2\times2\times2$,  $1\times2\times2$, $4\times2\times2$, and $2\times2\times1$ supercells for the \g-, \al-, \bo-, and B19-phases, respectively.

\section{Results}

\subsection{Site preference for stoichiometric phases}\label{sec:site}

The site preference energy, $E_s$, is defined as \cite{Wolf1996-ms}
\begin{equation}
  E^\xi_s(X)=E^\xi_{\mathrm{Ti}_{m-1}X\mathrm{Al}_n}-E^\xi_{\mathrm{Ti}_m X\mathrm{Al}_{n-1}}-E_{\mathrm{Al}}+E_{\mathrm{Ti}}\ ,\label{eq:site}
\end{equation}
where $E^\xi_{\mathrm{Ti}_{m-1}X\mathrm{Al}_n}$ and $E^\xi_{\mathrm{Ti}_mX\mathrm{Al}_{n-1}}$ are total energies of supercells of the phase $\xi$ with one Ti and Al atom, respectively, replaced with one $X$ atom, while $E_{\mathrm{Al}}$ and $E_{\mathrm{Ti}}$ are total energies of Al and Ti atom, respectively, in their stable crystal structure (Al:fcc and Ti:hcp). As explicitly derived by, e.g., \citet{Erschbaumer1993-ih}, and detailed in Sec.~\ref{sec:DeltaEf}, $E_s^\xi(X)$ expresses the difference in energy of formation of substitutional defects $\ce{Ti}\to X$ ($X$ occupies a Ti-site) and $\ce{Al}\to X$ ($X$ occupies an Al-site). $E_s^\xi(X)<0$ suggests that $\ce{Ti}\to X$ is an energetically preferred defect to $\ce{Al}\to X$, hence it is a indicator for a preferential occupation of the Ti sublattice sites (and hence yielding Al-rich compositions, a terminology used throughout this manuscript), and vice versa.

\begin{figure}
  \centering
  \includegraphics[width=\figWidth]{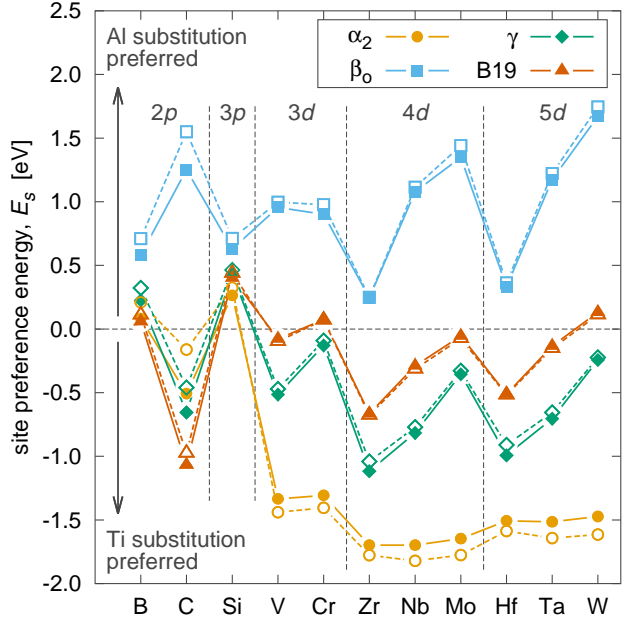}
  \caption{Calculated site-preference energy, $E_s$, for \al-\ce{Ti3Al}, \bo-,\g-, and B19-TiAl phases for selected group IIIA, IVA, and early TM alloying elements. The solid and dashed lines correspond to GGA and LDA calculations, respectively.}\label{fig:site}
\end{figure}

Figure~\ref{fig:site} summarises $E_s^\xi(X)$ for a large set of group IIIA and IVA elements (labelled as $2p$ and $3p$) as well as early transition metals (labelled as $3d$, $4d$, and $5d$). It is clear that there is only a negligible difference between the predictions based on GGA-PW91 and LDA xc potentials, both suggesting qualitatively the same behaviour. The two phases, \g and \al, appearing in the equilibrium Ti-Al binary phase diagram \cite{Ohnuma2000-ee} show a strong preference for transition metal (TM) elements occupying Ti sites, while they preferred to sit on the Al sublattice in the mechanically unstable \bo-phase \cite{Holec2015-gs}. The B19-phase mostly exhibits preference for the Ti substitution, however, this is by far the weakest among the four here considered phases, and for, e.g., the group VIB elements Cr and W turns even into a slight preference for the Al substitution. It is also worth noting that, irrespective of the phase, $E_s$ gets more positive (less negative) with increasing valence electron configuration, hence suggesting an inclination towards Al substitution when going from VIB to IVB group elements within 1 row of the periodic table of elements.

Interestingly, B and Si prefer to sit on the Al sublattice (Ti-rich compositions) for all 4 here investigated phases, while C gives the same behaviour as the TM elements. Nevertheless, since B and C are relatively small atoms, they may actually occupy interstitial instead of substitutional positions \cite{Klein2015-tz}. Our test calculations for these cases showed that indeed $E_f$ of TiAl with B and C atoms placed in the (tetragonal or octahedral) interstitial positions is lower than for the substitutional defects (see Tab.~\ref{tab:Ef}). In particular, Ti-rich octahedral configuration (i.e., an interstitial site bounded by an octahedron with vertices being 4 Ti and 2 Al atoms) gives lower energy of formation for both elements and all 4 phases than the Al-rich octahedral configuration (which does not even exist for the \al-phase) and the tetrahedral configuration. The latter is moreover unstable in the \al-, \bo- and B19-phases, as the interstitial atom moves into the neighbouring Ti-rich octahedral position during the relaxation of atomic positions. Furthermore, Ti-pure octahedral site (6 Ti nearest neighbours) exists in the \al-phase, and is energetically preferred over the Ti-rich configuration (4 Ti + 2 Al). The strong preference for the Ti-rich tetrahedral interstitial position is in agreement with previous literature reports, e.g., see \cite{oct1, oct2}.

\begin{table}
\centering
\caption{Energies of formation of the 4 TiAl phases studied here with B and C atoms in interstitial and substitutional positions. The second header row denotes the nearest neighbourhood configuration of interstitial atoms and type or substitution for the substitution atoms, respectively (see text for explanation).}\label{tab:Ef}
\begin{ruledtabular}
\begin{tabular}{lcccccc}
  & \multicolumn{3}{c}{octahedral} & \multicolumn{1}{c}{tetrahedral} & \multicolumn{2}{c}{substitutional} \\
  & 6 Ti & 2 Al + 4 Ti & 4 Al + 2 Ti & 2 Al + 2 Ti & $\text{Ti}\to X$ & $\text{Al}\to X$ \\
  \hline  
  B\\
  \al-\ce{Ti3Al} & $-0.2773$ & $-0.2551$ & (1) & (2) & $-0.2062$ & $-0.2124$ \\
  B19-TiAl & (1) & $-0.3737$ & $-0.2931$ & (2) & $-0.2785$ & $-0.2935$ \\
  \bo-TiAl & (1) & $-0.3783$ & $-0.3047$ & (2) & $-0.1802$ & $-0.1977$ \\
   \g-TiAl & (1) & $-0.3936$ & $-0.3578$ & $-0.3375$ & $-0.3209$ & $-0.3446$ \\
  \\
  C\\
  \al-\ce{Ti3Al} & $-0.3199$ & $-0.2836$ & (1) & (2) & $-0.1476$ & $-0.1555$ \\
  B19-TiAl & (1) & $-0.3961$ & $-0.2869$ & (2) & $-0.2233$ & $-0.2288$ \\
  \bo-TiAl & (1) & $-0.4183$ & $-0.3262$ & (2) & $-0.1317$ & $-0.1293$ \\
  \g-TiAl & (1) & $-0.4170$ & $-0.3733$ & $-0.3199$ & $-0.2644$ & $-0.2844$   
\end{tabular}
\end{ruledtabular}
\begin{tabbing}
  (1) \= position does not exist \\
  (2) \> \parbox{\columnwidth}{\raggedright atom relaxes into the 2 Al + 4 Ti octahedral configuration}
\end{tabbing}
\end{table}

\subsection{Effect of alloy composition}

\begin{figure}
  \centering
  \includegraphics[width=\figWidth]{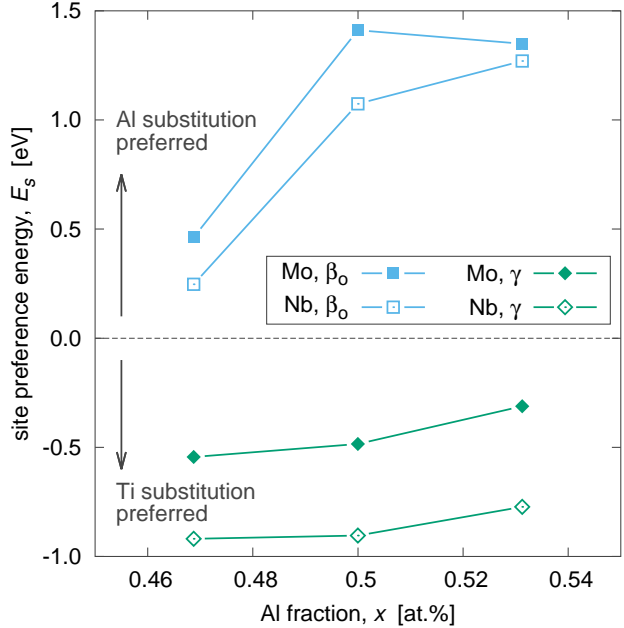}
  \caption{Dependence of the site preference energy, $E_s^\xi(X)$, on the Al content of the parent binary phase, evaluated for $X=$Mo and Nb and two phases $\xi=$\bo and \g.}\label{fig:off}
\end{figure}

Compositions of application-related alloys deviate from the perfect stoichiometry, i.e. from Ti$_{0.5}$Al$_{0.5}$. We have therefore used a similar approach as in Sec.~\ref{sec:site} to evaluate the impact of the Al content on the site preference for two technologically important elements, Nb and Mo, which are known as strong \bo-phase stabilisers \cite{Appel2011-tf, Clemens2013-jd}.

For example, the structural model for Al-rich composition was obtained by replacing one Ti atom with one Al. In such a supercell, all possible replacements $\ce{Al} \to X$ and $\ce{Ti} \to X$ were considered. The corresponding energies were averaged and the resulting values were inserted to Eq.~\ref{eq:site}. \change{Strictly speaking, only the lowest energy configuration should be taken into account at $0\uu{K}$. With increasing temperature, probability of realising the higher energy states becomes $\exp(\beta E_i)/Z$, where $E_i$ is the energy of a state (configuration) $i$, $\beta=1/k_BT$ is the thermodynamic $\beta$, and $Z=\sum_i \exp(\beta E_i)$ is the partition sum. It follows, that the simple averaging as used here corresponds to the high-temperature limit. However, the error in the estimation of the site preference energy caused by this simplification is in the range of only a few meV already at $T=1\uu{K}$ and exponentially converges to the high-temperature limit, and therefore the below discussed trends are relevant for any practical implications.}

The thus obtained compositional dependences are shown in Fig.~\ref{fig:off}. Within the investigated range of $\approx\pm3\uu{at.\%}$ off-stoichiometry, the site preference remains the same as for the pure structures. While for the Al-rich compositions the preference for Ti-substitution in the \g-phase decreases, the same trend is observed for the \bo-phase for Al-lean compositions and Al-substitution. This can be intuitively understood by the fact, that, e.g., Al-rich compositions imply fewer Ti-sites remain available for the substitution. In such case, Ti-substitution does mean further deterioration of the Ti sublattice, while Al-substitution happens statistically most often on the Al-sublattice which is still intact.

\subsection{Alloying impact on alloy stability}\label{sec:DeltaEf}

Energy of formation, $E_f$, is a measure of stability of a certain compound with respect to its building elements. In the following we evaluate the change of the energy of formation, $\Delta E_f$, when a third element is alloyed into the parent phase. If, for example, we consider Ti$_{m-1}X$Al$_n$ in the $\xi$ phase ($\xi=$\al, \bo, \g, or B19), then its corresponding $\Delta E_f^\xi(X)\big|_{\mathrm{Al-rich}}$ reads
\begin{equation}
  \Delta E_f^\xi(X)\big|_{\mathrm{Al-rich}} = E_f^\xi(\ce{Ti}_{m-1}X\ce{Al}_n)-E_f^\xi(\ce{Ti}_m\ce{Al}_n)\ .\label{eq:DeltaEf}
\end{equation}
The energies of formation of TiAl$+X$ and TiAl, respectively, in the above equation are given per atom, and are evaluated as
\begin{widetext}
\begin{align}
  E_f^\xi(\ce{Ti}_{m-1}X\ce{Al}_n) &= \frac1{m+n}\left[E^\xi_{\ce{Ti}_{m-1}X\ce{Al}_n}-\Big((m-1)E_\ce{Ti}+E_X+nE_\ce{Al}\Big)\right]\ ,\label{eq:EfTiAlX}\\
  E_f^\xi(\ce{Ti}_m\ce{Al}_n) &= \frac1{m+n}\left[E^\xi_{\ce{Ti}_m\ce{Al}_n}-\Big(mE_\ce{Ti}+nE_\ce{Al}\Big)\right]\ .\label{eq:EfTiAl}
\end{align}
$E_X$ is the total energy (per atom) of the element $X$ in its stable bulk phase, the rest of the quantities were defined in Sec.~\ref{sec:site}. Putting Eqs.~\ref{eq:EfTiAlX} and \ref{eq:EfTiAl} into Eq.~\ref{eq:DeltaEf} yields
\begin{equation}
  (m+n)\Delta E_f^\xi(X)\big|_{\mathrm{Al-rich}} = E^\xi_{\ce{Ti}_{m-1}X\ce{Al}_n}-\Big(E^\xi_{\ce{Ti}_m\ce{Al}_n}-E_\ce{Ti}+E_X\Big)\ .\label{eq:DeltaEfalt}
\end{equation}
\end{widetext}
The right-hand side of last equation expresses the formation energy for a point substitutional defect $\ce{Ti}\to X$. Consequently, the substitutional defect formation energy and the alloying-induced change of the alloy energy of formation are, up to a scaling factor $(m+n)$, the same quantities. Analogous formula can be derived also for $\Delta E_f^\xi(X)\big|_{\mathrm{Ti-rich}}$. It then follows 
\begin{equation}
 (m+n)\left[\Delta E_f^\xi(X)\big|_{\mathrm{Al-rich}} - \Delta E_f^\xi(X)\big|_{\mathrm{Ti-rich}}\right] = E_s\ ,
\end{equation}
hence difference between the alloying-induced changes in energies of formation for Al-rich and Ti-rich compositions shows the same trend as the site preference energy.

\begin{figure*}
  \begin{subfigure}[t]{0.49\textwidth}
    \caption{Ti-rich}\label{fig:Ef_Al}
    \includegraphics[width=\figWidth]{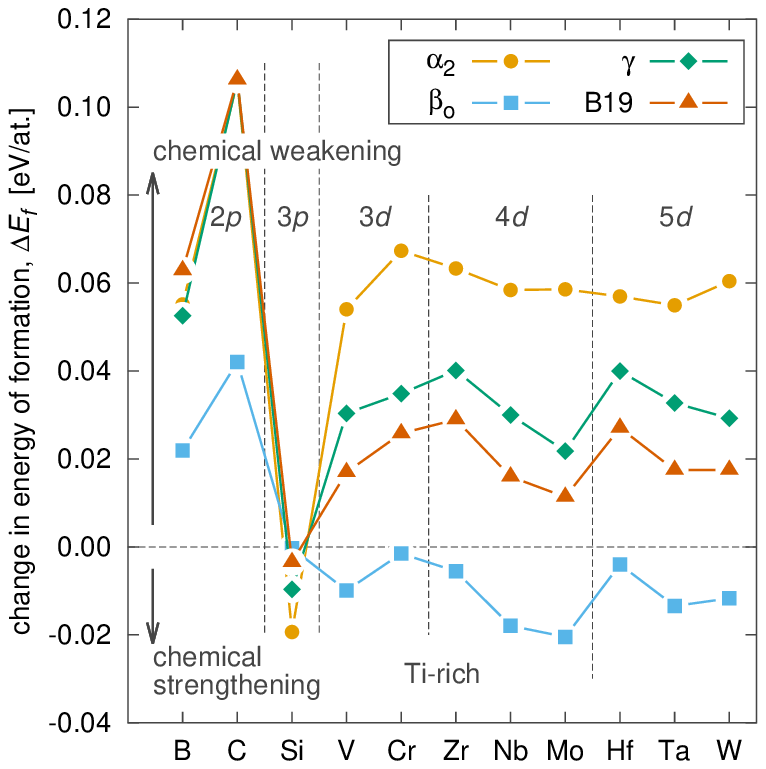}
  \end{subfigure}
  \begin{subfigure}[t]{0.49\textwidth}
    \caption{Al-rich}\label{fig:Ef_Ti}
    \includegraphics[width=\figWidth]{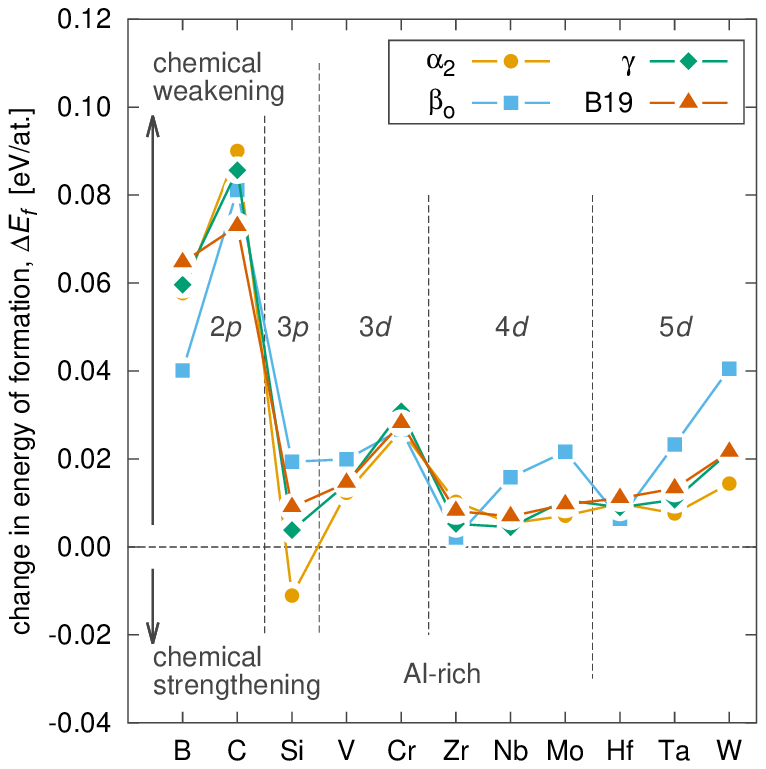}
  \end{subfigure}
  \caption{Alloying-related change in the energy of formation,$E_f$, for \al-, \bo-, \g-, and B19-phases. Panels (\subref{fig:Ef_Al}) and (\subref{fig:Ef_Ti}) correspond to the Ti-rich and Al-rich compositions, respectively, after the ternary addition $X$.}\label{fig:DeltaEf}
\end{figure*}

The calculated $\Delta E_f$ is shown in Fig.~\ref{fig:DeltaEf} where positive or negative values describe situations in which alloying yields destabilising or stabilising, respectively, of the individual TiAl phases. Substitution of $2p$ elements, B and C, leads to a strong destabilisation of all 4 phases for both, Ti-rich and Al-rich compositions. Si, on the other hand, stabilises the \al-, \g-, and B19-phases for the Ti-rich compositions. It is also the only element exhibiting chemical strengthening for the Al-rich situations, namely for the \al-phase. Hence Si is expected to be strongly inclined to be present in the \al-\ce{Ti3Al}. This prediction has been confirmed by a recent experimental study~\cite{Klein2016-au}.

Regarding the TM elements, they cause chemical destabilisation of the \al-, \g-, and B19-phases for both Ti-rich (stronger) and Al-rich (weaker) compositions. Interestingly, their alloying impact is similar for all 4 phases (i.e., including the \bo-TiAl) on the Al-rich side. Finally, TM elements chemically stabilise the \bo-TiAl for Ti-rich compositions, an effect discussed in detail for the TiAl+Mo alloy system in our recent publication \cite{Holec2015-gs}. This result also reflects the strong preference of TM elements to substitute for Al in the \bo-phase (cf.~Sec.~\ref{sec:site} and Fig.~\ref{fig:site}).

\section{Discussion}

\subsection{Current results in the context of the existing literature}

Our predictions are well in line with the extensive literature which, however, covers only a limited number of alloying elements and mostly focuses on the \g-phase. V, Zr, Nb, Mo, Ta, and W have been reported to occupy Ti-sites in the \g-phase \cite{Woodward1998-wl, Song2002-cy, Song2000-oz, Zhou2010-ev, Khowash1993-gy, Nandy1990-ih, Erschbaumer1993-ih, Wolf1996-ms, Jinlong1992-pg, Benedek2005-oq}, in agreement with our predictions. Some controversy remains in the case of Cr: While most recent reports suggest preference for Al-substitution \cite{Song2000-oz, Song2002-cy, Dang2007-ct, Erschbaumer1993-ih, Wolf1996-ms, Hao1999-ex}, others suggested preference for Ti-substitution \cite{Jinlong1992-pg, Khowash1993-gy}. Since the value of $E_s^\gamma(\ce{Cr})$ is actually very close to $0\uu{eV}$, we suggest that this controversy may be related to different calculation methods (linearised muffin-tin orbital approach vs. full potential augmented plane wave method, LDA vs. GGA, spin-polarised vs. spin-non-polarised). A similar situation exists for Mo and W in the \g-phase, which have been suggested to occupy both, Al and Ti sublattices \cite{Nandy1990-ih, Woodward1998-wl}. Moreover, works dealing with more elements confirm the dependence of $E_s^\xi(X)$ on the valence electron concentration \cite{Song2002-cy, Song2000-oz, Dang2007-ct, Wolf1996-ms, Woodward1998-wl, Hao1999-ex, Benedek2005-oq}. In addition, using electronic structure calculations not based on DFT, \citet{Hao1999-ex} concluded that Hf, Ta, Zr, Nb, Mo, V, and Cr strongly prefer to substitute for Ti atoms in the \al-\ce{Ti3Al}, which is in perfect agreement with the present predictions as well as with other DFT-based results by \citet{Benedek2005-oq}. Finally, the one report on Mo preference in the \bo-TiAl+Mo alloy again agrees with our predictions for Mo occupying Al sites \cite{Singh2008-ek}. 

The experimental observation of the site occupation of the sublattices of ordered structures is a difficult task. Nevertheless, we attempt to discuss this on the basis of very accurate phase resolved chemical data as obtained from atom probe tomography (APT) which provides simultaneously information on the type and position of individual atoms. Several researchers have published data on the chemical composition of the \g-, \al-, and \bo-phases \cite{Gerstl2004-tv, Scheu2009-xa, Klein2015-tz, Larson1997-ft}. These consistently show that the \g-phase contains between 49 and $50\uu{at.\%}$ of Ti and between 44 and $45\uu{at.\%}$ of Al. The remainder is made up by the other alloying elements, which are mainly refractory metals such as Nb and Mo. The apparent disagreement with our predictions may be explained by reference to an earlier study, which claimed that the refractory elements in fact occupy the Ti sublattice, and in turn transfer some of the Ti atoms to the Al sublattice \cite{Boll2007-av}. We discuss this scenario more in detail in Sec.~\ref{sec:anti-site}.

In case of the \al-phase, the Ti concentrations presented in literature range from 56 to $59\uu{at.\%}$, while the Al compositional range is very broad between 32 and $40\uu{at.\%}$ \cite{Gerstl2004-tv, Scheu2009-xa, Larson1997-ft}. The reduced Ti concentrations in comparison to the ideal stoichiometry of \ce{Ti3Al} suggest that the refractory metals again occupy the Ti sublattice, in agreement to our calculations. Apparently, in multi-component alloys the stability range of this phase may be altered in a way that a significant amount of Al is accommodated on Ti sites, which, however, goes beyond the present study.

The \bo-phase is characterized by a Ti concentration between 53 and $56\uu{at.\%}$ and an Al concentration between 33 and $36\uu{at.\%}$. As most of the refractory metals are $\beta$-stabilizing elements, large amounts of these elements are typically observed in this phase \cite{Klein2015-tz, Song2014-my}. In agreement with our predictions, these elements are averred to occupy the Al sublattice, thereby replacing Al and resulting in the reduced Al content of this phase. Unfortunately, there is no experimental data available regarding the chemical composition of the B19-phase, however, the previous discussion underpins the applicability of \textit{ab initio} calculations for the prediction of the site occupation of ordered structures. 

Based on the above discussion we suggest that the present predictions are sufficiently backed up by the existing theoretical as well as experimental data, and therefore bring novel (for \bo- and B19-phases as well as for some elements, e.g., Si) and coherent information. 

\subsection{Chemical and elastic contributions}

\begin{figure*}
  \begin{subfigure}[t]{0.49\textwidth}
    \caption{Ti-rich, chemical strengthening}\label{fig:Echem_Al}
    \includegraphics[width=\figWidth]{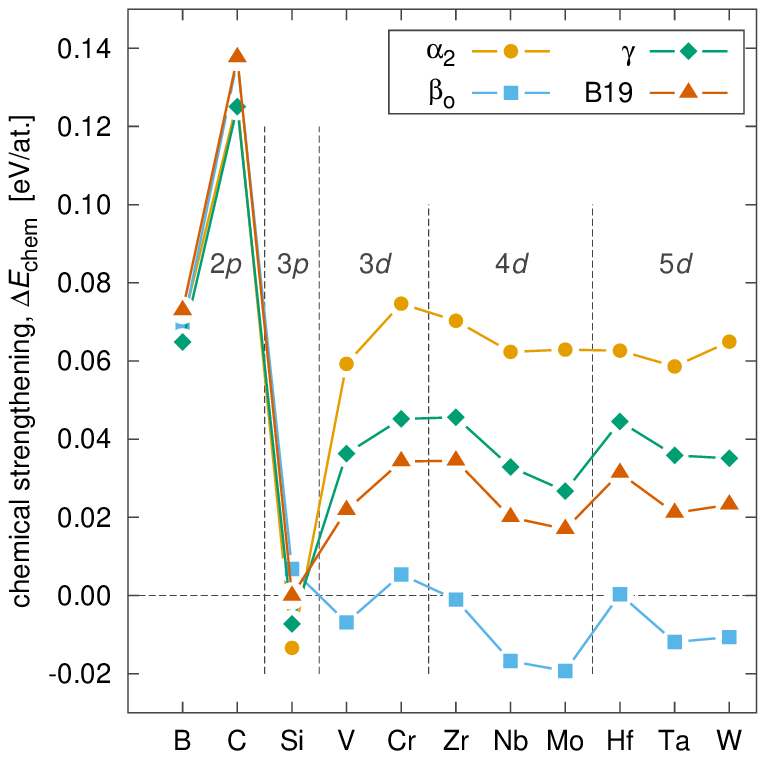}
    \end{subfigure}
  \begin{subfigure}[t]{0.49\textwidth}
    \caption{Ti-rich, elastic relaxations}\label{fig:Eelast_Al}
    \includegraphics[width=\figWidth]{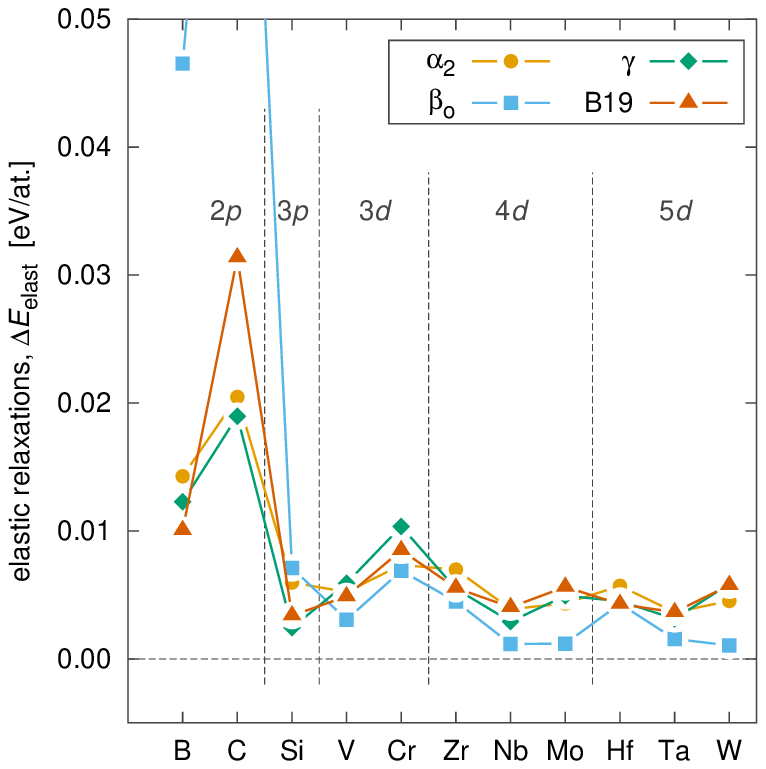}
  \end{subfigure}
  \caption{Chemical (\subref{fig:Echem_Al}) and elastic (\subref{fig:Eelast_Al}) contributions to the alloying-induced changes in the energy of formation, $\Delta E_f$, as shown in Fig.~\ref{fig:Ef_Al} for Ti-rich compositions. Note the different scales of the energy axes in both graphs.}\label{fig:DeltaEfContributions}  
\end{figure*}

When a foreign atom is substituted into the crystal, it causes two perturbations. One concerns its different chemistry, and hence different electronic and bonding environment as compared with the parent phase. The second disruption is related to the different size of the atom, and the thus induced local displacements of the neighbouring atoms. In the following, the former effect is termed chemical weakening/strengthening ($\Delta E_{\mathrm{chem}}$) while the latter contribution are elastic relaxations ($\Delta E_{\mathrm{elast}}$).

A practical decomposition of $\Delta E_f=\Delta E_{\mathrm{chem}}-\Delta E_{\mathrm{elast}}$ is done as follows~\cite{Holec2013-yt}: $\Delta E_{\mathrm{chem}}$ is the difference between $E_f$ of the perfect structure and the structure with one substituted atom (Al in Fig.~\ref{fig:DeltaEfContributions}). The atom positions in the TiAl+$X$ system are kept fixed to those of the perfect system so to account only for the changed chemistry. In the second step, the atomic positions are allowed to relax, i.e., to lower the total energy of the system, while the chemistry remains unchanged.

Figure~\ref{fig:DeltaEfContributions} shows these two contributions for all 4 phases and all alloying elements considered here for the Ti-rich compositions ($\ce{Al}\to X$). It can be concluded that the major contribution to $\Delta E_f$ (Fig.~\ref{fig:Ef_Al}) originates from the chemical changes induced by alloying, as (i) Fig.~\ref{fig:Echem_Al} largely resembles the trends in Fig.~\ref{fig:Ef_Al}, and (ii) $|\Delta E_{\mathrm{chem}}|$ (Fig.~\ref{fig:Echem_Al}) is in most cases by factor 5 and more larger than $\Delta E_{\mathrm{elast}}$ (Fig.~\ref{fig:Eelast_Al}). Importantly, the amount of the elastic energy is very similar for all phases, unlike $\Delta E_{\mathrm{chem}}$, hence suggesting that $\Delta E_{\mathrm{elast}}$ is predominantly determined by the size of the foreign atom rather than the parent phase (its structure and bonding).

It is also worth noting the predicted contributions for the \bo-TiAl. Interestingly, the chemical contribution $\Delta E_{\mathrm{chem}}$ is practically zero for the Ti-isovalent elements Zr and Hf, which, on the other hand, possess one of the largest values of $\Delta E_{\mathrm{elast}}$ for the \bo among the TM elements. Opposite situation is predicted for $4d$ and $5d$ group VB and IVB elements (Nb, Mo, Ta, and W) where the chemical strengthening term is dominant while the elastic relaxations are negligible. It can be therefore concluded, that while all TM elements investigated here yielded increased stability for the Ti-rich \bo-TiAl phase (Fig.~\ref{fig:Ef_Al}), the dominant reason for this strengthening is different for the group IVB elements on the one hand (different atomic sizes), and the group VB and IVB elements on the other hand (different valency).

The by far largest values for both contributions were obtained for B and C. This is related to their small sizes ($E_{\mathrm{elast}}$) and hence their preferential occupation of the interstitial instead of substitutional positions (see Sec.~\ref{sec:site} and Tab.~\ref{tab:Ef}).

Lastly, we note that in the case of Al-rich compositions (Fig.~\ref{fig:Ef_Ti}) the general trends (e.g., $|\Delta E_{\mathrm{elast}}|\ll\Delta E_{\mathrm{chem}}$) remain similar to those discussed above.

\subsection{Alloying-induced changes in electronic structure of TiAl+Mo}

In order to further underline the critical role of alloying elements on the local electronic and bonding structure, we have chosen the $\beta$-phase stabilising element Mo for a closer analysis. It shows a strong site preference for the occupation of Al sites in the \bo-phase, while it only slightly favours the Ti sublattice in the \g-phase (Fig.~\ref{fig:site}). 

\begin{figure*}
  \begin{subfigure}{0.49\textwidth}
    \caption{\bo-TiAl}\label{fig:DOS_beta}
    \includegraphics[width=\figWidth]{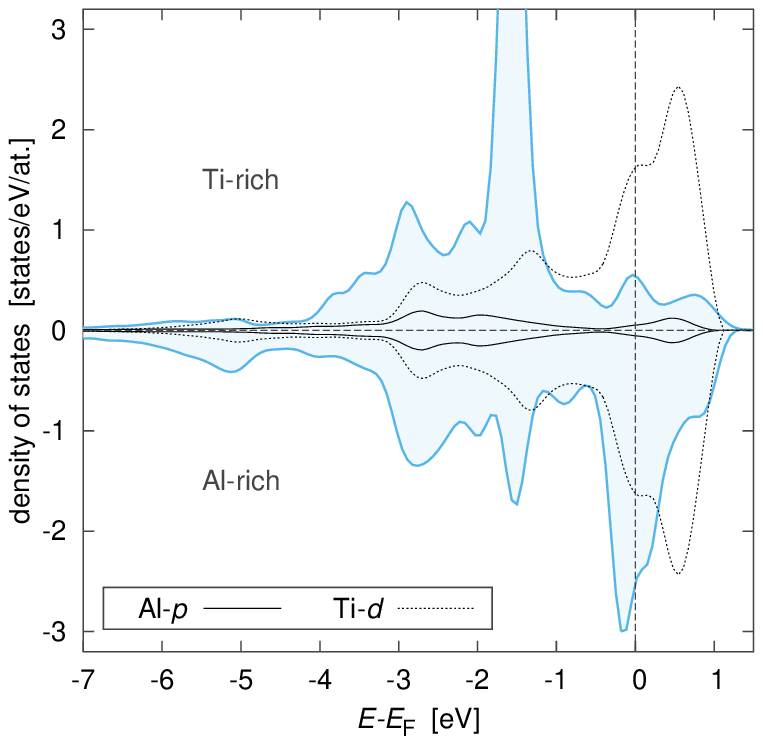}
  \end{subfigure}
  \begin{subfigure}{0.49\textwidth}
    \caption{\g-TiAl}\label{fig:DOS_gamma}
    \includegraphics[width=\figWidth]{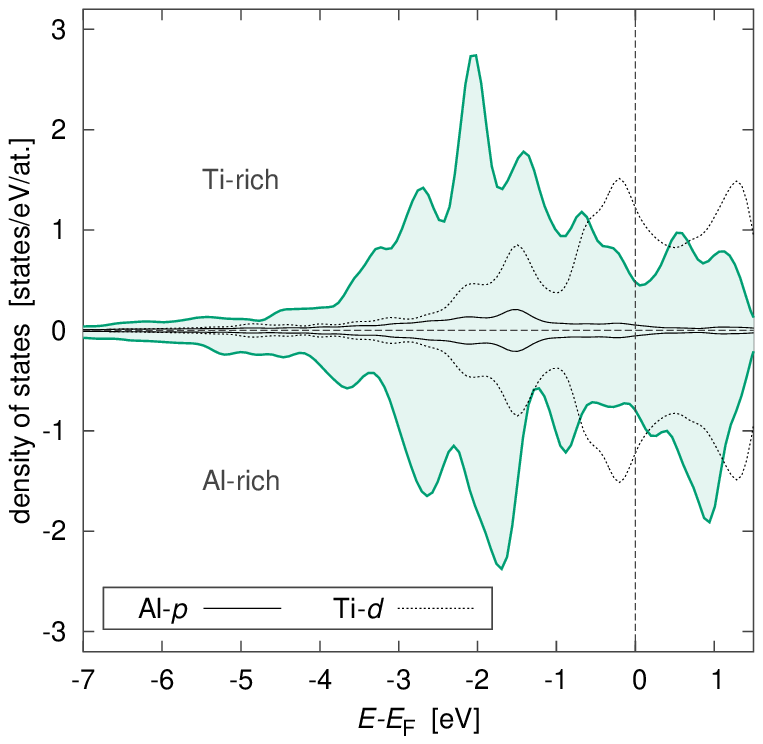}
  \end{subfigure}
  \caption{Local density of states corresponding to the Mo site in (\subref{fig:DOS_beta}) the \bo-TiAl and (\subref{fig:DOS_gamma}) the \g-TiAl Ti-rich (upper) and Al-rich (lower) solid solutions are shown with the thick solid colour lines. The thin black solid (Al-$p$) and dotted (Ti-$d$) lines represent projected density of states in respective perfect stoichiometric phases for comparison.}\label{fig:electronic}
\end{figure*}

Local density of electronic states (LDOS) on the Mo site is shown in Fig.~\ref{fig:electronic} for the Ti-rich and Al-rich compositions for \bo- (Fig.~\ref{fig:DOS_beta}) and \g-TiAl (Fig.~\ref{fig:DOS_gamma}) together with the Ti$-d$ and Al$-d$ partial projected density of electronic states (PDOS) of the corresponding parent phases. The large peak at the Fermi level ($E_F$) for the Al-rich \bo-phase suggests its lower stability in comparison with the small LDOS at $E_F$ in the case of the Ti-rich compositions~\cite{Ravindran1997-kc}. This is the underlying origin for the preferential substitution $\ce{Al}\to\ce{Mo}$ in the \bo-phase, as the phase stability is directly linked with the alloying-induced changes of $E_f$ (Fig.~\ref{fig:DeltaEf}) which determine the site preference energy, $E_s$. Similarly, the LDOS for the \g-phase (Fig.~\ref{fig:DOS_gamma}) shows no significant differences at $E_F$, hence similar stabilities of both Ti$_{n-1}X$Al$_n$ and Ti$_nX$Al$_{n-1}$, as also the corresponding $E_s$ suggests only a slight preference for the Ti substitution (cf.~Fig.~\ref{fig:site}).

\subsection{Anti-site defects}\label{sec:anti-site}

\begin{figure*}
  \begin{subfigure}[t]{0.49\textwidth}
    \caption{Ti-rich}\label{fig:Ef_Al_as}
    \includegraphics[width=\figWidth]{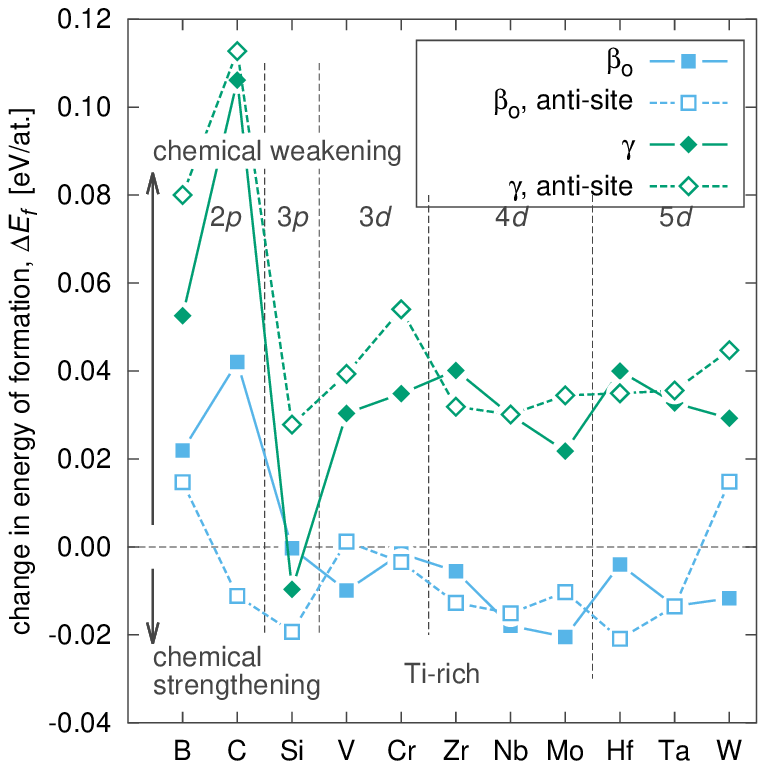}
  \end{subfigure}
  \begin{subfigure}[t]{0.49\textwidth}
    \caption{Al-rich}\label{fig:Ef_Ti_as}
    \includegraphics[width=\figWidth]{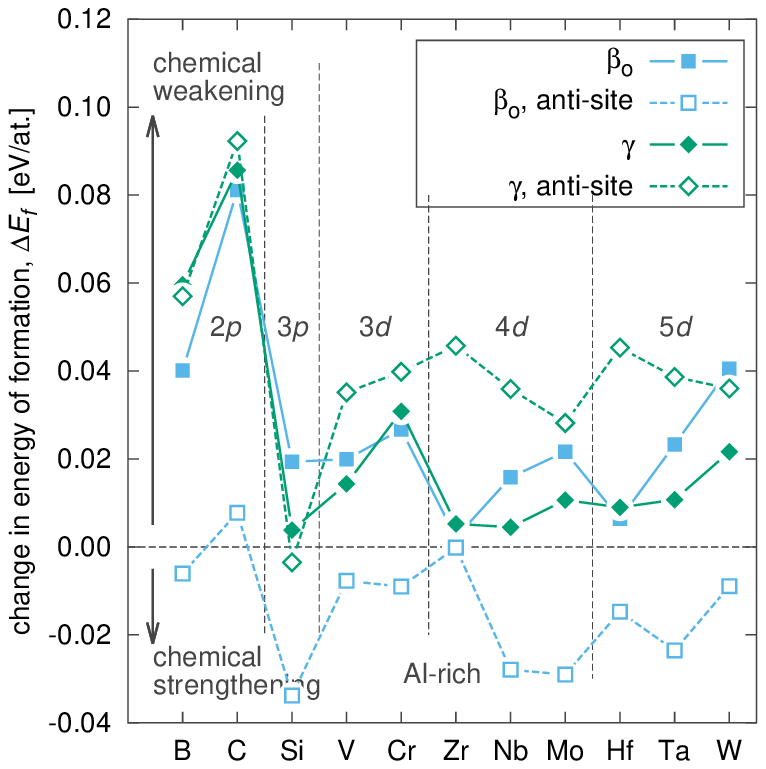}
  \end{subfigure}
  \caption{Comparison of the change in the energy of formation $|\Delta E_f|$ induced by a ternary addition as a simple substitutional defect (solid lines) or a double defect combining substitutional and anti-site defects (dashed lines) for the \bo- (blue) and \g-phases (green) resulting in a \subref{fig:Ef_Al_as} Ti-rich and \subref{fig:Ef_Ti_as} Al-rich compositions.}\label{fig:DeltaEf_as}
\end{figure*}

The Ti-rich composition, for instance, may be obtained also by a combination of a substitution $\ce{Ti} \to X$ and formation of an anti-site defect $\ce{Al}\to\ce{Ti}$. To assess the energetics of this complex scheme with respect to a simple substitution $\ce{Al} \to X$ (yielding the same Ti-rich composition), we have considered an anti-site defect in the nearest neighbour distance from the substitutional defect in the \bo- and \g-phases. The results in term of the induced change of the energy of formation are shown in Fig.~\ref{fig:DeltaEf_as}.

Considering the Ti-rich compositions (Fig.~\ref{fig:Ef_Al_as}), the combination of a substitutional atom and an anti-site is not hugely different from a single substitutional atom on the opposite sublattice for either \bo- or \g-TiAl. Inspecting the early TM element, however, it is noticeable that while the group VIB elements (Cr, Mo, W) prefer the simple substitution $\ce{Al} \to X$ in the \g-phase, the group VB elements (V, Nb, Ta) exhibit smaller preference or are ambivalent, and finally the group IVB elements (Zr, Hf) prefer the $\ce{Ti} \to X$ and $\ce{Al}\to\ce{Ti}$ double defect. A similar result is predicted also for the \bo-phase with an overall somewhat stronger preference to the simple substitution. Interestingly, the $\ce{Ti} \to X$ and $\ce{Al}\to\ce{Ti}$ mechanism is strongly preferred for C and Si in the \bo-phase. We note, however, that C, which is known as a strong $\alpha_2$-phase stabiliser, has almost no solubility on the $\beta$-phase~\cite{Klein2016-au}.

On the contrary, a very different behaviour is predicted for the Al-rich compositions. While in the \g-phase, a simple substitution $\ce{Ti} \to X$ is the energetically preferred mechanism for all early TM over the double defect ($\ce{Al} \to X$ combined with the Ti$\to$Al anti-site), and there is no preference for B, C and Si, the \bo-phase is contrarily suggested to adopt the double defect mechanism. Even more importantly, $\Delta E_f^\xi(X)\big|_{\mathrm{Ti-rich}}$ for a simple substitution (single-site defect) for the Ti-rich compositions is less negative than $\Delta E_f^\xi(X)\big|_{\mathrm{Al-rich}}$ for the double-site defect for the Al-rich compositions. Nevertheless, the site occupation for ternary alloying elements $X$ remains unchanged, as the Al-rich double-site defect incorporates $X$ on the Al sublattice, in agreement with the conclusions based on Fig.~\ref{fig:site}. A more detailed discussion including, e.g., dependence of the distance between the substitutional and anti-site defect, or off-stoichiometry effects, however, goes beyond the present study.

\Change{Finally, let us note that substitution employing anti-sites was also discussed by \citet{Jiang2008-sn}, who evaluated $E_{X}^{\mathrm{Al}\to\mathrm{Ti}}$ expressing the energy change upon a ternary element $X$ being shifted from the Al to the Ti sublattice with a simultaneous creation of a Ti anti-site (in the Ti-rich case). Such an approach had been defined earlier by \citet{Ruban1997-fr}. On the one hand, it has a clear advantage of avoiding the single element energies $E_{\mathrm{Al}}$ and $E_{\mathrm{Ti}}$ in the evaluation, and allows one to estimate whether site preference is weak or strong~\cite{Jiang2008-sn}. On the other hand, it assumes a limit of non-interacting substitutional site and anti-site, in contrast with the our approach in which both defects are present in one simulation supercell at the same time. Importantly, both approaches yield a consistent story. For example, \citet{Jiang2008-sn} predicted that while all Ta, Zr, or Nb atoms occupied Ti sites in \g-Ti$_{0.47}$Al$_{0.51}X_{0.02}$ (Al-rich), this happened only for about $80\,\%$ of Mo substitutional atoms (the rest sit on the Al sublattice). These predictions were confirmed by experimental measurements~\cite{Hao1999-ex}. Our calculations lead to similar conclusions, as (i) all these elements preferentially occupy the Ti sublattice (cf. Fig.~\ref{fig:site}, and Fig.~\ref{fig:off} for compositional dependence in the case of Mo and Nb), and (ii) the anti-site occupation is energetically strongly unfavourable compared with the direct substitution in the case of Ta, Zr, and Nb (yielding almost zero $\Delta E_f$), while the energy difference between the anti-site and the direct implementation is the smallest for Mo (cf. Fig.~\ref{fig:Ef_Ti_as}). Therefore, generation of anti-sites at elevated temperatures proceeds easiest in the Mo case.}

\section{Conclusions}

\textit{Ab initio} quantum-mechanical calculations were used to predict which sublattice is preferentially occupied by ternary additions to the ordered binary TiAl phases, i.e., \g, \al, \bo, and B19. While the Al substitution is strongly favoured for the \bo-phase (B2 structure), early transition metal elements (V, Cr, Zr, Nb, Mo, Hf, Ta, and W) will occupy Ti sites in the \g- (L$1_0$) and \al-phases (D$0_{19}$). Early TM elements prefer Ti sublattice also in the B19-TiAl, however, with increasing valence electron concentration the Al substitution becomes energetically favourable. B and Si exhibit preference for the Al sublattice for all four phases. Substitutional C prefers Ti-sites in the \al, \g, and B19 phases, and would prefer Al site in the \bo-TiAl. However, we note that it has almost no solubility in latter phase. Nevertheless, interstitial incorporation of B and C into Ti-rich octahedral positions yields lower energy of formation than the above mentioned substitutional positions.

The trends in the preferred sublattice occupation have been correlated with the alloying-induced changes of the energy of formation, $\Delta E_f$, i.e. the stability of the solid solution. Here, apart from Si in all phases, and all TM elements in \bo-TiAl, all other alloying elements in all phases result in their destabilisation. Further analysis revealed that the chemical part (change in the electronic structure) of $\Delta E_f$ is dominant over the elastic strains related to the different atomic size than the parent species. These changes have been also linked qualitatively to the changes in the electronic structure, namely the density of electronic states at the Fermi level.

Finally, we have provided an evidence that an incorporation of the ternary element on one sublattice, combined with an anti-site defect, is in some cases (e.g., Al-rich compositions of the \bo-TiAl) preferred over a direct substitution on the other sublattice. 


%

\end{document}